\documentclass[aps,showpacs, twocolumn, amsmath,amssymb]{revtex4}
\begin{document}
\title{ {\Large {\bf Scalar sextet  in the 331 model with right-handed neutrinos}}}
\author{Nguyen Anh Ky}
\affiliation{Institute of Physics and Electronics, 10 Dao Tan, Hanoi, Vietnam}
\author{Nguyen Thi Hong Van}
\affiliation{Institute of Physics and Electronics, 10 Dao Tan, Hanoi, Vietnam}
\begin{abstract}
A Higgs sextet is introduced in order to generate Dirac and
Majorana neutrino masses in the 331 model with right-handed
neutrinos. As will be seen, the present sextet introduction leads
to a rich neutrino mass structure. The smallness of neutrino
masses can be achieved via, for example, a seesaw limit. The fact
that the masses of the charged leptons are not effected by their
new Yukawa couplings to the sextet is convenient for generating
small neutrino masses.
\end{abstract}
\pacs{12.10.Dm, 12.60.Fr, 14.60.Pq}
\maketitle
\section{\label{sec:level1} Introduction}

  The 331 model \cite{pp,fram, fhpp,flt} based
on the gauge group $SU(3)_C\otimes SU(3)_L\otimes U(1)$, is one of
the interesting extensions of the standard model (SM). Firstly, it
is stated that the model guarantees the number of the families to
be three. Secondly, the Peccei--Quinn symmetry \cite{pq} -- a
solution of the strong CP problem, naturally occurs in the 331
model \cite{pal}. The third reason making this model interesting
is that one of the quark families transforms under the gauge group
differently from the other two. This could naturally explain
unbalancing heavy top quarks in the fermion mass hierarchy,
deviations of $A_b$ from the SM prediction, etc. Recent analyzes
indicated that signals of new particles in this model, i.e.,
bileptons \cite{dion,a3m} and exotic quarks \cite{das}, may be
observed at the Tevatron and the LHC. The fact that the 331 model
predicts not very high new mass scales, at the order of a few
TeV's only \cite{ng}, may help us to solve in a near  future some
hard problems like the $g-2$ problem \cite{a3m}, neutrino
oscillations and masses (see, for example,
\cite{tj}--\cite{gusso}) , etc..\\

  The 331 model has several versions.
Two of them are the minimal version \cite{pp,fram,fhpp} (in which
no new lepton is introduced and all leptons of each generation are
grouped in one triplet), and the version with right-handed
neutrinos (RHN's) \cite {flt}. Each of the two versions has
specific feactures but at the beginning they both were introduced
with a Higgs sector consisting of three scalar $SU(3)_L$-triplets
(or shortly, triplets) only. Later, a Higgs $SU(3)_L$-sextet (or
shortly, a sextet) described by a symmetric tensor $S_{ij}=S_{ji},
~ i,j =1,2,3,$ was added to the minimal version in order to
properly generate all charged lepton masses \cite{fhpp}. In the
version with RHN's,  so far, however, a Higg sextet has not been
needed as the three Higgs triplets have been enough for generating
masses for all charged leptons and quarks. As far as neutrinos are
concerned, we can make them remaining massless by choosing an
appropriate configuration of vacuum expectation values (VEV's) of
the Higgs sectors. The problem related to fermion masses seems
quite simple here. Both the theoretical motivations and the
experimental results in the last time \cite{ska} -- \cite{wp},
however, give more and more evidences supporting the idea of
massive neutrinos which may be of a Dirac or a Majorana type. So
far, many theoretical mechanisms for generating neutrino masses
have been suggested (see, for example, \cite{smir} --
\cite{bipet87} and references therein). In the models with massive
Dirac, not Majorana, neutrinos, a certain combination of the
lepton numbers can be a conserved quantity, while an existence of
massive Majorana neutrinos violates it. So, the question of
massive neutrinos of  which type is directly related  to a
fundamental question of symmetries in particle physics. In the 331
models there have been several ways neutrinos to get masses.\\

  The minimal 331 model can generate Majorana masses which could be
tree-level (if a non-zero VEV of the sextet component
$S_{00}\equiv \sigma_1^0$ is allowed) \cite{tj} or radiatively
induced \cite{kitabay,okamoto}. There are also other mechanisms
for generating neutrino masses for this version of the 331 model,
for example, in \cite{mpp2} the sextet is replaced by a neutral
scalar singlet and a dimension-seven effective operator is
introduced. Experimental and practical data show that the
neutrinos if massive have very tine masses (only a few eV's or
less). In the frameworks of the minimal 331 model some attempts
for explaining the smallness of neutrinos masses have been made
(see, for eaxmple, \cite{tj} -- \cite{mpp2}). As far as the 331
model with RHN's is concerned, investigations on neutrinos masses
are much poorer here. In this model, the neutrinos (more
precisely, two of them) can gain Dirac masses if a definite one of
the three Higgs triplets develops a non-zero VEV but at the
present, as stated in \cite{flt}, it is not known how to get small
neutrino masses. Additionally, there is no explicit evidence for
the Majorana neutrino masses to be excluded in advance from
consideration when the lepton number has no real meaning in both
versions of the 331 model (as a lepton and its antiparticle are
simultaneosly components of one and the same multiplet) and when
neutrinoless double beta ($(\beta\beta)_{0\nu}$) decays (which are
sensitive to the existence of Majorana neutrinos and violate the
total lepton number by two uinits) are still considered as
possible processes \cite{sde} -- \cite{2-beta-bi}. Moreover, the
presence of the right-handed neutrinos in the theory is a good
reason for considering neutrino masses of Majorana type. In this
paper we show that Majorana neutrino masses can be introduced in
the 331 model with RHN's by adding a scalar sextet to the Higgs
sector. Beside Majorana neutrino masses, this modified model
allows Dirac neutrino masses too. The smallness of the neutrino
masses can be achieved by taking, for example, a seesaw limit. In
general, as will be seen, the sextet introduction suggests a rich
structure of the neutrino masses. The present paper is organized
as follows.\\

  In the next section we briefly recall some necessary elements of
two versions of the 331 model, namely the minimal version and the
original version with RHN's in which the Higgs sector contains
only three scalar triplets. A scalar sextet is introduced to this
sector in the Section 3 where we show how neutrino  masses can be
generated after the sextet introduction. The conclusion and some
comments are given in the last section, Section 4.

\section{ \label{sec:level1} Some elements of the 331 models}
\subsection{\label{sec:level2} The minimal version}

  Let us start with the minimal version of the 331 model
\cite{pp,fram,fhpp}. The leptons of all generations transform
under the gauge group $SU(3)_C\otimes SU(3)_L\otimes U(1)$ (or the
331-gauge group, for short) in one and the same way as  follows
\begin{equation}f^a_L = \left(\begin{array}{c}
\nu_L^a\\[2mm]
l_L^a\\[2mm]
(l_R^a)^c
\end{array}\right) \sim ( 1, 3, 0), ~~ l_R^c\equiv (l_R)^c,
\label{lepton}
\end{equation}
where $a=1,2,3$, is a generation index.
The Higgs sector of the minimal 331 model consists of three triplets
and one sextet which ensure masses for all fermions
in the model via symmetry breaking which could follow the order
\begin{equation}
\begin{array}{c}
SU(3)_C \otimes SU(3)_L \otimes U(1)_N\\[2mm]
\langle \chi \rangle \downarrow\\[2mm]
SU(3)_C \otimes SU(2)_L \otimes U(1)_Y\\[2mm]
\langle \rho \rangle, \langle \eta \rangle
\downarrow \langle S \rangle \\[2mm]
SU(3)_C \otimes U(1)_Q, \end{array}\label{sb}\end{equation}\\
where $\langle \chi \rangle$, $\langle \rho \rangle$, $\langle
\eta \rangle$ and $\langle S \rangle$ are VEV's of the Higgs
fields $\chi$, $\rho$, $\eta$ and $S$, rescpectively (see
\cite{fhpp,m331higgs} for more details). As far as the quark
sector is concerned, one of the quark generations transforms
differently from the other two which in turn transform in one and
the same way under the 331-gauge group.
\subsection{\label{sec:level2} The version with right-handed neutrinos}

  In  the original version of the 331 model with RHN's proposed in \cite{flt} a lepton multiplet of
each generation, obtained from the corresponding one in (\ref{lepton}) by replacing
the third component $l_R$ with a right-handed neutrino $\nu_R$, tranforms under
the 331-gauge group as
\begin{equation*}f^a_L =
\left(\begin{array}{c} \nu_L^a\\[2mm] l_L^a\\[2mm]
(\nu_R^c)^a
\end{array}\right) \sim (1, 3, -1/3), \  l_R^a \sim (1, 1, -1),
\end{equation*}
where $a$ is a generation index. In our opinion, the right-handed
neurtino on the third component of $f_L$ could be in general
different from the anti-neutrino $\nu_R$ - the anti-particle of
the neutrino $\nu$, staying on the first component of $f_L$.  To
avoid any confusion, instead of $\nu_R^c$ (used in \cite{flt}) we
suggest \cite{comments} another notation, say $N_R^c$, for the
third component of $f_L$:
\begin{equation}f^a_L =
\left(\begin{array}{c} \nu_L^a\\[2mm] l_L^a\\[2mm]
(N_R^c)^a
\end{array}\right) \sim (1, 3, -1/3), \  l_R^a \sim (1, 1, -1).
\end{equation}
Here, as in the minimal version, two of the quark generations
transfrom in one and the same way, while the remaining one
transforms differently (as the quark sector is not considered in
this paper, its explicit stucture and transformation are not given here).\\

  The Higgs sector of the original version with RHN's \cite{flt}
consists of three scalar triplets
\begin{eqnarray}
\eta&=& \left( \begin{array}{c}
\eta_1^0 \\[2mm] \eta_2^- \\[2mm] \eta_3^0
\end{array} \right)
\sim (1,3,-1/3),~~
\rho= \left( \begin{array}{c}
\rho_1^+ \\[2mm] \rho_2^0 \\[2mm] \rho_3^{+}
\end{array} \right)
\sim(1,3,2/3),\nonumber\\
\chi&=& \left( \begin{array}{c}
\chi_1^0 \\[2mm] \chi_2^{-} \\[2mm] \chi_3^0
\end{array} \right)
\sim (1,3,-1/3).
\end{eqnarray}
The Yukawa couplings in this case
\begin{eqnarray}
{\cal L}^{\chi}_{yuk} &=& \lambda_1 \bar Q_{1L} u^{'}_{1R} \chi
+ \lambda_{2ij} \bar Q_{iL} d^{'}_{jR} \chi^{\dagger} + H.c.,
\label{ykw}
\nonumber \\[2mm]
{\cal L}_{Yuk}^{\rho} &=& \lambda_{1a} \bar Q_{1L} d_{aR} \rho
+ \lambda_{2ia} \bar Q_{iL} u_{aR} \rho^{\dagger} \nonumber\\&&
+ G_{ab}\bar f^a_L(f^b_L)^c \rho^{\dagger}
+ G^{'}_{ab} \bar f_L^a e_R^b \rho + H.c.,
\\[2mm]
{\cal L}^{\eta}_{yuk} &=& \lambda_{3a} \bar Q_{1L} u_{aR} \eta +
\lambda_{4ia} \bar Q_{iL} d_{aR} \eta^{\dagger} + H.c.,
\nonumber
\end{eqnarray}
where $a,b=1,2,3$, $i=2,3$, can ensure (via the LHS symmetry breaking
scheme in (\ref{sb})) masses  for all quarks and charged leptons as
well as Dirac
masses for two of  the neutrinos \cite{flt}. This model, however, cannot
explain the smallness of neutrino masses and does not generate (at least,
at the tree-level) a Majorana neutrino mass which, as discussed above,
is by no reason to be exluded in advance from consideration. This
problem may be solved by introducing a scalar sextet to the  model.
\section{\label{sec:level1} Sextet and neutrino masses in the 331 model
with RHN's}

   A neutrino mass (at the tree level) can be generated by coupling an
appropariate Higgs boson to $\bar f_L(f_L)^c$. As an $SU(3)_L$-tensor
the latter is a (tensor) product ${\bf 3}^{*}\otimes  {\bf 3}^{*}$ of
two $SU(3)_L$--anti-triplet $\bar f_L$ and $(f_L)^c$, consequently, it
can be decomposed into a direct sum of a triplet {\bf 3} (the
anti-symmetric part of the tensor) and an anti-sextet {\bf 6}$^{*}$
(the symmetric part of the tensor):
$${\bf 3}^{*}\otimes {\bf 3}^{*}= {\bf 3}\oplus {\bf 6}^{*}.$$
To constitute an $SU(3)_L$-invariant quantity we can contract
$\bar f_L(f_L)^c$ with an anti-triplet ${\bf 3}^{*}$ and/or a
sextet {\bf 6}. The term $G_{ab}\bar f^a_L(f^b_L)^c\rho^{\dagger}$
in the Yukawa Lagrangians (\ref{ykw}) is a contraction of the
first kind. Here, three $SU(3)_L$-indeces are antisymmetrized by
contracting with the anti-symmetric $SU(3)_L$-tensor
$\epsilon_{ijk}$, i.e., the rank-two tensor $\bar f_L(f_L)^c$ is
anti-symmetrized (therefore, it transforms under $SU(3)_L$ as a
triplet {\bf 3}) and contracted with $\rho^{\dagger}$, an
anti-triplet ({\bf 3}$^{*}$). At $\langle \rho \rangle \neq 0$,
the above Yukawa coupling term can generate Dirac masses for two
of the three neutrinos (while the third  one remains massless)
\cite{flt}. This way for generating neutrino masses, however,
gives no indication for that the neutrino masses obtained are
small, and it excludes the Majorana neutrino masses which (as
explained above) might be important. A scalar sextet added to the
Higgs sector may solve this questions.\\

   A scalar field tranforming under $SU(3)_L$ as a sextet {\bf 6} can be
described by a symmetric tensor which in the present case has the following
explicit form and 331-gauge transformation law
\begin{equation}
{\cal S}= \left( \begin{array}{ccc}
\tau_1^0 & ~ T_1^-/\sqrt{2} & ~ \tau_2^0 /\sqrt{2}\\[3mm]
T_1^- /\sqrt{2} & ~ T_2^{--} & ~ T_3^- \\[3mm]
\tau_2^0 /\sqrt{2}& ~ T_3^- & ~ \tau_3^0
\end{array} \right)
\sim (1,6,-2/3).
\end{equation}
A non-zero VEV of this sextet ${\cal S}$ coupled to the symmetric
part of $\bar f_L(f_L)^c$ could give rice to Dirac and/or Majorana
neutrino masses without effecting the masses of the charged
leptons. The Lagrangian term  corresponding to this Yukawa
coupling is
\begin{equation}
G^s_{ab}\bar f^a_L(f^b_L)^c{\cal S} + \mbox {H. c.},
\label{ykws}
\end{equation}
where $G^s_{ab}$ are new coupling constants; $a,b=1,2,3$, are generation indeces,
while the $SU(3)_L$-indeces are omitted. A general structure
of a VEV of ${\cal S}$ could be
\begin{equation}
\langle {\cal S}\rangle = \left( \begin{array}{ccc}
\omega_1 & ~ 0  & ~ \omega_2 /\sqrt{2}\\[3mm]
0 & ~ 0 & ~ 0 \\[3mm]
\omega_2 /\sqrt{2}& ~ 0 & ~ \omega_3
\end{array} \right),
\end{equation}
where $\omega_n$ are VEV's of the neutral sextet components $\tau^0_n$,
$n=1,2,3$. This VEV when inserted in (\ref{ykws}) leads to the mass
term \\
\begin{equation}
G^s_{ab}\left(\bar {\nu}_L^{~a}, ~ \bar l_L^{~a}, ~
(\bar N_R^{~c})^a\right)
\left( \begin{array}{ccc}
\omega_1 & ~ 0  & ~ \omega_2 /\sqrt{2}\\[3mm]
0 & ~ 0 & ~ 0 \\[3mm]
\omega_2 /\sqrt{2}& ~ 0 & ~ \omega_3
\end{array} \right)
\left(\begin{array}{c} (\nu_L^{~c})^b\\[2mm] (l^c_L)^b\\[2mm]
N_R^{~b}
\end{array}\right),
\end{equation}
which in the neutrino subspace has the form\\
\begin{equation}
G^s_{ab}\left(\bar {\nu}_L^{~a}, ~ (\bar N_R^{~c})^a\right)
\left( \begin{array}{ccc}
\omega_1 & ~ \omega_2 /\sqrt{2}\\[3mm]
\omega_2 /\sqrt{2}& ~ \omega_3
\end{array} \right)
\left(\begin{array}{c} (\nu_L^{~c})^b\\[2mm] N_R^{~b}
\end{array}\right).
\end{equation}
The latter is nothing but the familiar Dirac--Majorana mass term \\
\begin{equation}
 {1\over 2} \left(\bar {\nu}_L, ~ \bar N_R^{~c}\right)
\left( \begin{array}{ccc}
{\bf m}_T & ~ {\bf m}_D\\[3mm]
{\bf m}_D & ~ {\bf m}_S
\end{array} \right)
\left(\begin{array}{c} \nu_L^{~c}\\[2mm] N_R
\end{array}\right),
\label{m-term}
\end{equation}
where the generation indeces are omitted and ~ ${\bf m}_{T,D,S}$ ~ are $3\times 3$
matrices with the following elements
\begin{eqnarray}
({\bf m}_T)_{ab} &=&2G^s_{ab}~\omega_1, ~~
({\bf m}_D)_{ab} = \sqrt{2}G^s_{ab}~\omega_2,
\nonumber\\[2mm]
({\bf m}_S)_{ab} &=& 2G^s_{ab}~\omega_3.
\end{eqnarray}
An analysis of a mass term of this kind is well known. For example,
at the seesaw limit
\begin{equation}
{\bf m}_T\approx 0,~~  {\bf m}_S \gg {\bf m}_D
\end{equation}
or equivalently,
\begin{equation}
\omega_1\approx 0,~~  \omega_3 \gg \omega_2
\label{seesaw}
\end{equation}
we get two eigen mass matrices (generation--mixing, in general)
\begin{equation}
{\bf m}_1 = ({\bf m}_D)^T({\bf m}_S)^{-1}{\bf m}_D, ~~ {\bf m}_2 = {\bf m}_S.
\end{equation}
The condition (\ref{seesaw}) could be accepted in some circumstance as
$\omega_3$ characterizes the energy scale of bearking $SU(3)$ down to
$SU(2)$ and therefore it must be much bigger than $\omega_2$ and $\omega_1$
characterizing the scales of breaking $SU(2)$ and $U(1)$:
\begin{equation}
\omega_3 \gg \omega_2 \gg \omega_1.
\end{equation}

  Diagonalizing the matrix $G^s_{ab}$ and keepping (\ref{seesaw}) we
get for each eigenstate of $G^s_{ab}$, a diagonalized mass matrix,
\begin{eqnarray}
(m_T)_k &=& 2G^s_k~\omega_1,~~
(m_D)_k = 2G^s_k~\omega_2/\sqrt{2},\nonumber \\[2mm]
(m_S)_k &=& 2G^s_k~\omega_3,~~ k=1,2,3,\label{mmdiago}
\end{eqnarray}
leading to the Majorana neutrinos
\begin{eqnarray}
(n_{1L})_k&=& (\nu_L)_k - {(m_D)_k \over
(m_S)_k}(N_R^c)_k,\\[2mm]
(n_{2L})_k &=&  {(m_D)_k\over (m_S)_k}(\nu_L)_k + (N_R^c)_k,
\end{eqnarray}
with masses
\begin{eqnarray}
(m_1)_k &=& (m_D^2)_k/(m_S)_k \equiv G^s_k
{(\omega_2)^2 \over \omega_3},\\[2mm]
(m_2)_k &=& (m_S)_k \equiv 2G^s_k~\omega_3,
\label{majomass}
\end{eqnarray}
where $(M)_k$ is a diagonal element of a diagonalized matrix ${\bf
M}$. An estimation of these masses can be made by taking bounds of
the VEV's $\omega_i$ and the coupling constant $G^s$ via analyzing
different processes (decay modes and branching ratios) and an
appropriate scalar potential in the Higgs sector with
participation of the sextet. However, basing on the experimental
data, theoretical imagination or assumption and some trick (the
ratio $m_1/m_2=(\omega_2/\omega_3)^2$ does not depend on the
coupling constants $G$) we can avoid this, a bit long,
procedure.\\

If we can neglect the generation mixing we should have, for each
generation, Majorana neutrinos
\begin{eqnarray}
(n_{1L})_a&=& (\nu_L)_a - {(m_D)_a \over
(m_S)_a}(N_R^c)_a,\\[2mm]
(n_{2L})_a &=&  {(m  _D)_a\over (m_S)_a}(\nu_L)_a + (N_R^c)_a,
\end{eqnarray}
with masses
\begin{eqnarray}
(m_1)_a &=& (m_D^2)_a/(m_S)_a \equiv G^s_a {(\omega_2)^2
\over \omega_3},\\[2mm]
(m_2)_a &=& (m_S)_a \equiv 2G^s_a~\omega_3,
\end{eqnarray}
respectively. If the magnitude of $m_D$ is in the order of those
of the charged lepton masses, i.e., $(m_D)_1 \sim m_e$,
$(m_D)_2\sim m_\mu$, $(m_D)_3\sim m_\tau$, while $m_S$ is around 1
TeV or bigger \cite{a3m,ng}, we get the following estimation on
the masses of neutrinos $\nu_e$, $\nu_\mu$ and $\nu_\tau$ :
\begin{equation}
m_{\nu_e}< 0.25 \mbox{~eV}, ~ m_{\nu_\mu} < 11.2 \mbox{~KeV}, ~
m_{\nu_\tau}< 3.16 \mbox{~MeV}.
\label{nonmixngmass}
\end{equation}
It is some thing like the normal hierarchy of neutrino masses. A
popular theoretical assumption (for a seesaw model), however, says
that $m_S$ related to the mass of a right-handed neutrino could be
very large, with a magnitude near to that of $M_{GUT}$. A value of
$m_S$ between the range $10^{10}$ GeV $<m_S<10^{15}$ GeV can
ensure the neutrino masses very small even for relatively large
$m_D$ as accepted in some models (e.g., $m_D\sim 200$ GeV in
\cite{alta}).\\

  Let us return to Eqs. (\ref{mmdiago}) -- (\ref{majomass}) since the
generation mixing is already a real fact \cite{ska} --
\cite{lsnd}. According to the recent data \cite{hei-mos} --
\cite{wp} giving an upper bound of neutrino masses in the order of
1 eV (or less) we get a relation between the magnitudes of $m_D$
and those of $m_S$:
$${m_D^2\over m_S}= 10^{-9} \mbox{~GeV}$$
from which if we take the bound 300 GeV $<m_S <$ 1 TeV \cite{a3m,ng}
we should have
\begin{equation}
0.548 \mbox{~MeV} <m_D < 1 \mbox{~MeV}.
\end{equation}
The latter bound is around the electron mass.
As mentioned above the ratio $m_1/m_2$,
\begin{equation}
{m_1\over m_2}=\left({\omega_2\over \omega_3}\right)^2,
\end{equation}
is somewhat "universal", that is, it is independent from the
coupling constants but the ratio $\omega_2 /\omega_3$. Thus, a
ratio can be predicted if the other one is known.\\

In a standard way \cite{seesaw} -- \cite{bipet87} we can consider
other limits of the mass term (\ref{m-term}): the pure Majorana
limit ($m_D=0$), the Dirac limit ($m_T=m_S=0$) or the pseudo-Dirac
limit (when $m_T\ll m_D$ and $m_S\ll m_D$), etc. In the latter
case, there are two almost-degenerate Majorana neutrinos with a
very small mass spliting. For $m_S=0$, $m_T\ll m_D$ the masses are
$m_{\pm}=m_D\pm m_T/2. $ Another case is when both $m_D$ and $m_S$
(and/or $m_T$) are very small and comparable. This case leads to
non-degenerate Majorana neutrinos and, as in the pseudo-Dirac
case, to signicant ordinary-sterile ($\nu^a_L-(N^c)^a_L$) mixings
(to be or not to be confirmed by LSND). However, the latest
counted cases may not be compatible with the symmetry breaking
hierarchy of the model.

\section{\label{sec:level1} Conclusion}
We have introduced in the 331 model with RHN's a new Higgs field,
a sextet under $SU(3)_L$, with the hope to generate proper
neutrino masses. In the original version of the model \cite{flt},
the neutrinos, more precisely, two of them, can have Dirac masses
only but the presence of the right-handed neutrinos in the model
indicates that it is possible and reasonable to include Majorana
neutrino masses, especially, when the processes like
$(\beta\beta)_{0\nu}$ decays are still considered to be possible.
The sextet introduction, suggesting a rich neutrino mass
structure, may solve this problem and generate neutrino masses in
a right order. Next, following \cite{m331higgs} we can investigate
the mass spectra of this extended scalar sector of the 331 model
with RHN's (thus we need a notion of the lepton number for the
case and follows \cite{tj} to construct  a lepton-number-conserved
potential).\\

{\it Note added}: Some earlier versions with two neutrinos in a
lepton triplet of the 331 model were considered in \cite{add1,
add2,add3}. A Higgs sextet was introduced and discussed briefly in
\cite{add2,add3}. However, the corresponding neutrino mass term
was analyzed in other aspects, its seesaw limit, therefore, the
seesaw mechanism, was not considered and Majorana neutrinos and
masses were almost neglected in \cite{add2,add3} (due to the
circumstance in their epoch), unlike in the present paper.
\begin{acknowledgements}
We would like to thank the referee for drawing our attention to
the works \cite{add1,add2,add3} giving us interesting information.
One of us (N.A.K.) would like to thank S. Petcov and A. Smirnov
for discussions and H.N. Long for the cooperation in doing earlier
works on the 331 model which are useful for the present work. We
also would like to thank S. Randjbar-Daemi for kind hospitality at
the Abdus Salam International Centre for Theoretical Physics,
Trieste, Italy.

This work was supported in part by the National Research Program
for Natural Sciences of Vietnam under Grant No  410804.

\end{acknowledgements}


\begin{thebibliography}{99}
\bibitem{pp} F. Pisano and V. Pleitez, Phys. Rev. D {\bf 46}, 410 (1992).
\bibitem{fram} P. H. Frampton, Phys. Rev. Lett. {\bf 69}, 2889 (1992).
\bibitem{fhpp} R. Foot, O. F. Hernandez, F. Pisano and V. Pleitez, Phys.
Rev. D {\bf 47}, 4158 (1993).
\bibitem{flt} R. Foot, H. N. Long and Tuan A. Tran,
 Phys. Rev. D {\bf 50}, R43 (1994).
\bibitem{pq} R. D. Peccei and H. R. Quinn, Phys. Rev.
Lett. {\bf 38}, 1440 (1977);
Phys. Rev. D {\bf 16}, 1791 (1977).
\bibitem{pal} P. B. Pal, Phys. Rev. D {\bf 52}, 1659 (1995).
\bibitem{dion} B. Dion, T. Gregoire, D. London, L. Marleau
and H. Nadeau, Phys. Rev. D {\bf 59}, 075006 (1999).
\bibitem{a3m} Nguyen Anh Ky, Hoang Ngoc Long and Dang Van Soa,
Phys. Lett. {\bf B486}, 140 (2000); Nguyen Anh Ky and Hoang Ngoc
Long, {\it "The anomalous magnetic moment of muon: from the E821
experiment to bilepton masses"}, hep-ph/0103247.
\bibitem{das}P. Das, P. Jain, and D. W. McKay, Phys. Rev.
D {\bf 59}, 055011 (1999); Y. A. Coutinho, P. P. QueirozFilho
and M. D. Tonasse, Phys. Rev. D {\bf 60}, 115001 (1999).
\bibitem{ng}D. Ng, Phys. Rev. D {\bf 49}, 4805 (1994).
 \bibitem{tj}, M. B. Tully and G. C. Joshi,
Phys. Rev. D {\bf 64}, 011301(R) (2001).
\bibitem{okamoto} Y. Okamoto and M. Yasue, Phys. Lett.
{\bf B466}, 267 (1999).
\bibitem{kitabay} T. Kitabayashi and M. Yasue, Phys. Lett.
{\bf B508}, 85 (2001);  Phys. Rev. D {\bf 63}, 095002 (2001).
\bibitem{mpp1} J. C. Montero, C. A. deS. Pires and
V. Pleitez, Phys. Lett. {\bf B502}, 167 (2001).
 \bibitem{mpp2} J. C. Montero, C. A. deS. Pires and
V. Pleitez, Phys. Rev. D {\bf 65}, 093017 (2002).
\bibitem{gusso}A. Gusso, C. A. de S. Pires and P. S. Rodrigues
da Silva, Mod. Phys. Lett. {\bf A18}, 1849 (2003).
\bibitem{ska} Super-Kamiokande collaboration, Y. Fukuda {\it et al}.,
Phys. Rev. Lett. {\bf 81}, 1562 (1998); Phys. Lett. {\bf B436}, 33
(1998); Phys. Lett. {\bf B433}, 9 (1998).
\bibitem{soudan} Soudan 2 collaboration, W. W. Allison {\it et al}.,
Phys. Lett. {\bf B449}, 137 (1999).
\bibitem{sno} SNO collaboration, Q. R. Ahmad {\it et al}.,
Phys. Rev. Lett. {\bf 89}, 011301 (2002); {\it ibid.}, 011302
(2002); V. Barger et al., Phys. Lett. {\bf B537}, 179 (2002); A.
Baudyopadhyay et al., Phys. Lett. {\bf B540}, 14 (2002); J.
Barcall et al., JHEP 0207, 054 (2002); G. L. Fogli, E. Lisi, A.
Marrone, D. Montanino and A. Palazzo, Phys. Rev. D {\bf 66},
053010 (2002); B. Aharmim et al, nucl-ex/0502021.
\bibitem{lsnd} LSND collaboration, C. Athanassopoulos {\it et al},
Phys. Rev. Lett. {\bf 81}, 1774 (1998); Phys. Rev. C {\bf 58}, 2489 (1998).
\bibitem{sde} Yu. Zdesenko, Rev. Mod. Phys. {\bf 74}, 663 (2002).
\bibitem{2-beta-majo} C. Aalseth et al, Phys. Atom. Nucl. {\bf 67}, 2002 (2004).
\bibitem{2-beta-pet}S. Petcov, {\it "Theoretical Prospects of Neutrinoless Double Beta Decay"}, hep-ph/0504166.
\bibitem{2-beta-cho}  S. Choubey, W. Rodejohann, Phys. Rev. D {\bf 72}, 033016 (2005).
\bibitem{2-beta-331} A. G. Dias, A. Doff, C. A. deS. Pires and P. S. Rodrigues da
Silva, Phys. Rev. D {\bf 72}, 035006 (2005).
\bibitem{2-beta-bi} S. M. Bilenky, A. Faessler, T. Gutsche and F. Simkovic,
Phys.Rev. D {\bf 72}, 053015 (2005).
\bibitem{hei-mos} Heidelberg-Moscow collaboration,
H. Klapdor-Kleingrothaus et al,
Nulc. Phys. Proc. Suppl.  {\bf 110}, 364 (2002);
Phys. Lett. {\bf B586}, 198 (2004).
\bibitem{raff} G. G. Raffelt, {\it "Astrophysical and cosmological
neutrinos"}, hep-ph/0208024.
\bibitem{betarev} S. Elliott and P.Vogel, Ann. Rev. Nucl. Part. Sci. {\bf 52}, 115 (2002).
\bibitem{pdg} S. Eidelman et al, {\it "Review of particle physics"}
(Particle data group), Phys. Lett. {\bf 592}, 1 (2004), http://pdg.lbl.gov.
\bibitem{alta}  G. Altarelli and F. Feruglio, New  J. Phys. {6}, 106 (2004).
\bibitem{mainz} Ch. Kraus et al, Eur.Phys.J. {\bf C40}, 447 (2005).
\bibitem{smir} A. Yu. Smirnov, Int. J. Mod. Phys. {\bf A19},1180 (2004).
\bibitem{wp} R. N. Mohapatra et al, {\it "Theory of Neutrinos: A White Paper"},
hep-ph/0510213.
\bibitem{seesaw} M. Gell-Mann, P. Ramond and R. Slansky,
{\it "Complex spinors and unified theories}", in {\it
"Supergravity}", P. van Nieuwenhuizen  and D. Friedman, eds.,
North Holland, Amsterdam, 1979, p. 315; R. N. Mohapatra and G.
Senjanovi\'c, Phys. Rev. Lett. {\bf 44}, 912 (1980).
\bibitem{moha} R. N. Mohapatra and P. B. Pal, {\it "Massive neutrinos in
physics and astrophysics"}, World Scientific, Singapore 1998.
\bibitem{vogel} F. Boehm and P. Vogel, {\it "Physics of massive neutrinos"},
Cambridge university press, Cambridge 1992.
\bibitem{langac} P. Langacker, {\it "Neutrino physics"}, hep-ph/0506257.
\bibitem{smirnov96} A. Yu. Smirnov and M. Tanimoto, Phys. Rev. D {\bf 55},
1665 (1997); A. Yu. Smirnov, {\it "Alternatives to the seesaw mechanism"},
hep-ph/0411194.
\bibitem{bipet87} S. M. Bilenky and S. T. Petcov, Rev. Mod.
Phys.  {\bf 59}, 671  (1987); Erratum: {\it ibid.} {\bf 60}, 575 (1988);
{\it ibid.} {\bf 61}, 169 (1989).
\bibitem{m331higgs} Nguyen Tuan Anh, Nguyen Anh Ky and Hoang Ngoc Long,
Int. J. Mod. Phys. {\bf A15}, 283 (2000); {\it ibid.} {\bf A16}, 541 (2001).
\bibitem{comments} Nguyen Anh Ky, {\it "Comments related to reading
"Static quantities of the $W$ boson in the $SU_L(3)\times U_X(1)$ model
with right-handed neutrinos"}, hep-ph/0504147.
\bibitem{add1} M. Singer, J. W. F. Valle and J. Schechter, Phys. Rev. D {\bf 22},
738 (1980).
\bibitem{add2} J. W. F. Valle and M. Singer, Phys. Rev. D {\bf 28}, 540 (1983).
\bibitem{add3} J. C. Montero, F. Pisano and V. Pleitez, Phys. Rev. D {\bf 47}, 2918
(1993).

\end{thebibliography}
\end{document}